# THz meta-foil – a new photonic material

H.O. Moser,[1,*] L.K. Jian[1], H.S. Chen[2,3], M. Bahou[1], S.M.P. Kalaiselvi[1], S. Virasawmy[1], S.M. Maniam[1], X.X. Cheng[2], S.P. Heussler[1], Shahrain bin Mahmood[1], B.-I. Wu[2,3]

[1]*Singapore Synchrotron Light Source (SSLS), National University of Singapore (NUS), 5 Research Link, Singapore 117603*

[2] *The Electromagnetics Academy at Zhejiang University, Zhejiang University, Hangzhou 310058, China*

[3]*Research Laboratory of Electronics, Massachusetts Institute of Technology, Cambridge, Massachusetts 02139, USA*

*Corresponding author: moser@nus.edu.sg

**Seeing sharper [1-3] or becoming invisible [4-6] are visions strongly driving the development of THz metamaterials. Strings are a preferred architecture of metamaterials as they extend continuously along one dimension. Here, we demonstrate that laterally interconnecting strings by structural elements that are placed in oscillation nodes such as to not quench electromagnetic resonances enables manufacturing of self-supported free-standing all-metal metamaterials. Upright S-strings, interconnected by rods, form a space-grid which we call meta-foil. In this way, we introduce binding between the "atoms" of the metamaterial, thus doing away with conventional "frozen-in solutions" like matrix embedding or thin films on substrates. Meta-foils are locally stiff, yet globally flexible. Even bent to cylinders of 1 cm radius, they maintain their spectral response, thus becoming true metamaterials on curved surfaces. Exploiting UV/X-ray lithography and ultimately plastic moulding, meta-foils can be cost-effectively manufactured in large areas and quantities to serve as optical elements.**

Controlling and manipulating electromagnetic radiation by its interaction with materials is a cornerstone of optics. Over the past years, the study of materials that allow selecting their permittivity and permeability freely from positive to negative values has led to new concepts like the sub-wavelength-resolution imaging [1-3] or the invisibility cloaking [4-6]. Lately, even mimicking celestial mechanics in metamaterials has been proposed [7]. However, the development of practical metamaterials for the THz range up to the visible is lagging behind their theoretical study and envisaged applications. A practical metamaterial would be available in copious quantities and enable customized solutions, much like a foil, whereas the majority of present-day metamaterials is made by time-consuming primary pattern generation and involves dielectric substrates or matrices, which might substantially restrict their usefulness and applicability due to electric, mechanical, and thermal properties of dielectrics as well as their sensitivity to humidity and radiation degradation [8-14].

In earlier work, we mitigated such restrictions by micromanufacturing of free-standing metamaterials from metal-string arrays suspended in free space by plastic window-frames [15]. Made of gold, strings were S-shaped longitudinally. Aligned assembly of two chips of strings created a bi-layer chip in which two layers of S-strings, typically 1 to 10 µm apart, were opposed such as to form the well-known S-string resonator loops [16]. However, window-frames were still rigid and restricted the useful range of incidence angles.

Here, we demonstrate an all-metal approach where the conducting metal is the structural material simultaneously and no rigid window-frame is needed anymore. Individual S-strings are connected by transverse rods creating a space-grid that is self-supporting, locally stiff, but globally flexible. Connections are made between oscillation nodes of strings to minimize any influence on the resonance. Looking like a foil, such space-grid is baptized "meta-foil". Meta-foils can be tailor-made to virtually any shape, bent, and wrapped around objects to hide and shield them from electromagnetic radiation.

Figure 1 shows meta-foils and their geometric parameters. The plane of the meta-foil corresponds to the y-z plane. S-strings stand upright with leg "b" normal to that plane. Adjacent strings are shifted along z by (a-h)/2 with respect to each other. Strings are interconnected by rods that run along the y axis, are centered on leg "b", and repeated with 1S or 2S periods. A normally incident wave propagates in x-direction along leg "b", with electric and magnetic field vectors pointing along z and y axes, respectively. Inductive loops are formed by half an S in one row and the oppositely oriented half S in the adjacent row. Overlapping legs "b" form capacitances, loop areas determine the inductance. The time-varying magnetic field normal to the loops induces an electromotive force that drives a loop current.

Meta-foils are manufactured by means of 3-level lithography using either UV or X-rays depending on structure geometry. Figure 1 illustrates three layers. Middle and top layers need precise alignment of mask and substrate as well as re-deposition of a gold plating base. Bottom and top layers contain all conductors parallel to the y-z plane, the middle layer the "vias" and interconnecting rods that run along x- and y-directions, respectively. Top and bottom layers are related by a translation of (a-h)/2 along z. Hence, only two masks are needed, in principle, one for bottom and top layers, and one for the middle layer. However, for practical reasons, three masks are used (see Methods). To keep the resonance frequency within 10% of the nominal value, alignment accuracy must be <20% of the gap between S–strings which determines capacitance and resonance frequency. Omitting details, we note that this 3-level process is also suitable for plastic moulding, enabling cost-effective large-volume manufacturing of meta-foils [17].

Meta-foils presented here feature S-strings that are either equidistant and interconnected after one or two periods of "a" (1SE or 2SE), or grouped in pairs that are further apart than gap d (1SP or 2SP). For measured dimensions see Fig. 1. The spectral behavior of samples was characterized by Fourier transform interferometry in the far infrared between 2-14 THz. Beam spot size on sample was 1.5 mm at normal incidence, total beam divergence 60°. Figure 2 displays measured transmission spectra (top) of a 1SE foil versus frequency with the incidence angle α around the z axis running from 0°(9°)81° in comparison with simulated spectra (bottom). Peak positions and widths agree fairly well. MWS commercial software [18] was used for full-wave simulation. Two dominant peaks appear at 3.2 THz and

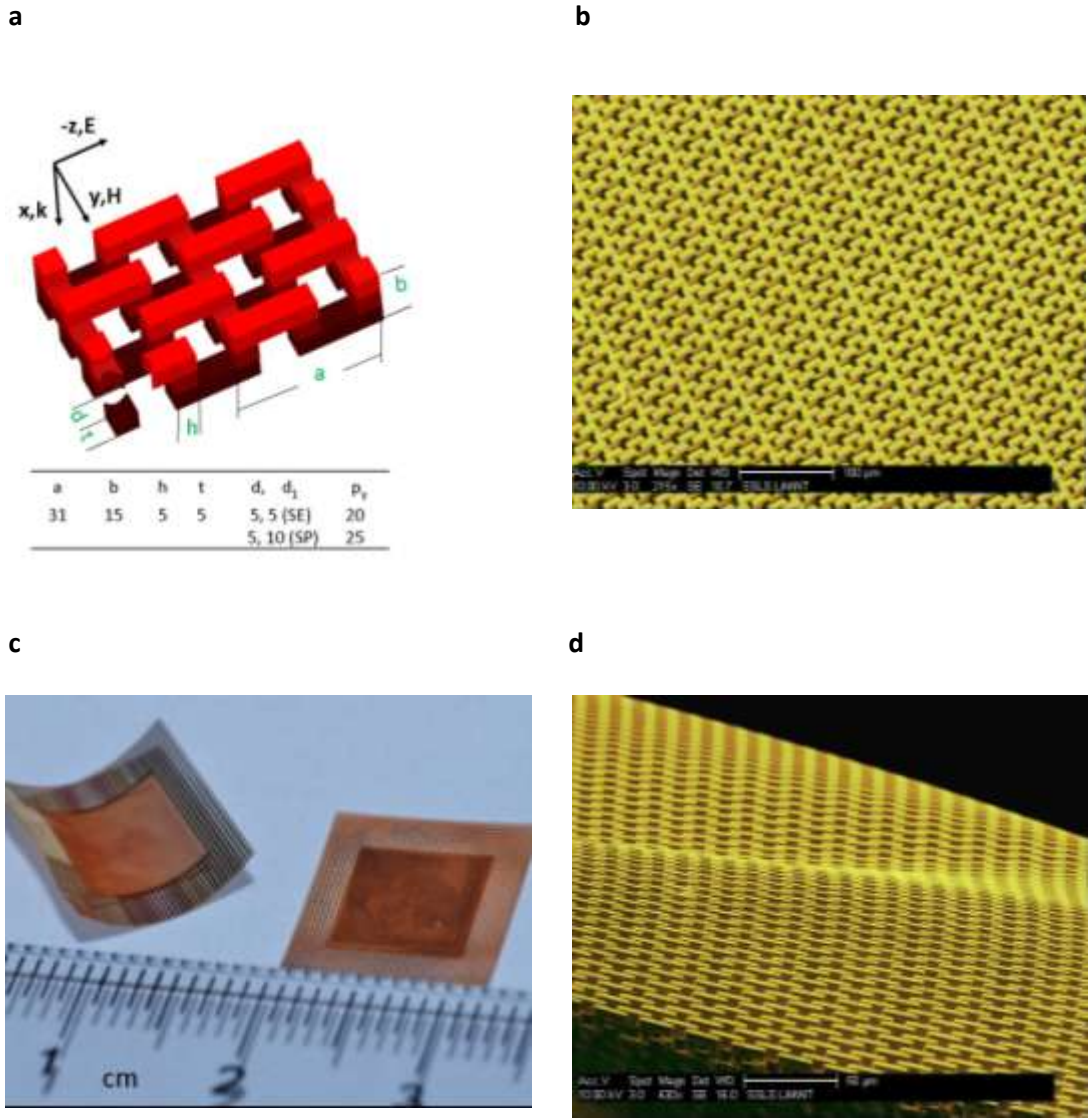

Fig. 1: The meta-foil. a) 3D schematic of a 1SE meta-foil showing lithography layers, coordinate frame, and parameter definition. Measured geometric parameters are given in the Table (unit μm). b) Scanning electron microscope (SEM) image of a 2SP Au meta-foil, scale bar 100 μm. c) Photo of flat and rolled Au meta-foils 8×7×0.015 mm$^3$ (L×W×H) in size. d) SEM image of a warped 1SP Au meta-foil, scale bar 50 μm.

6.8 THz. From parameter and index retrieval calculations [19] (Fig. 3), the peak at 3.2 THz (λ=94 μm) is assigned the well-known left-handed resonance of the fig-8 loop in S-strings [16]. Its wavelength-structure-size ratio of λ/b = 94/15 = 6.26 indicates a reasonable effective-medium approximation. The figure-of-merit $FOM = |Re(n)/Im(n)|$ is 5.17 at 3.25 THz and 4.13 at 3.0 THz from simulation. The peak at 6.8 THz (λ=44 μm) is a right-handed electrical resonance of one half S acting as an antenna between coupling capacitors or interconnecting rods as nodes. Here, the length of the shortest such resonator is a/2+h = (15.5+5) μm = 20.5 μm, in 7%-agreement with λ/2 = 22 μm.

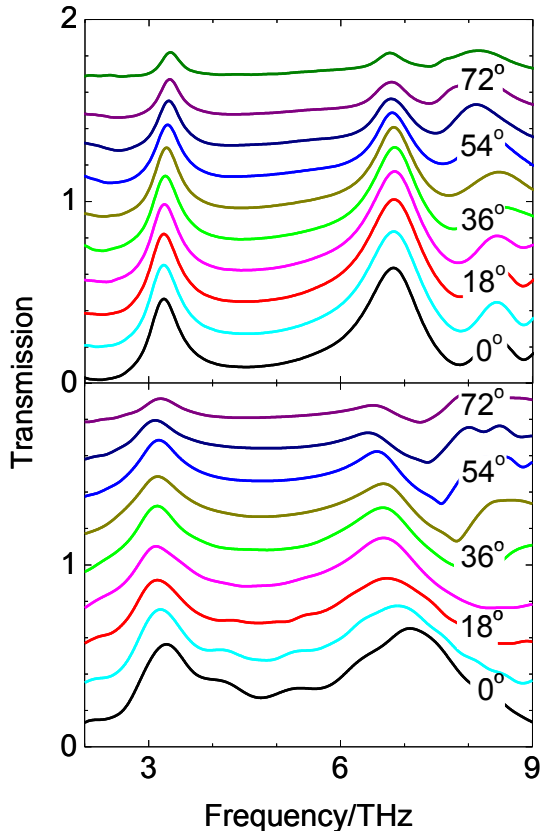

Fig. 2: Spectral transmission of meta-foils. Measured (top) and simulated (bottom) transmission spectra of a 1SE sample with the incidence angle $\alpha$ as a parameter varying from 0°(9°)81°, the electric field pointing along the z axis. Spectra are vertically shifted with respect to each other for clarity.

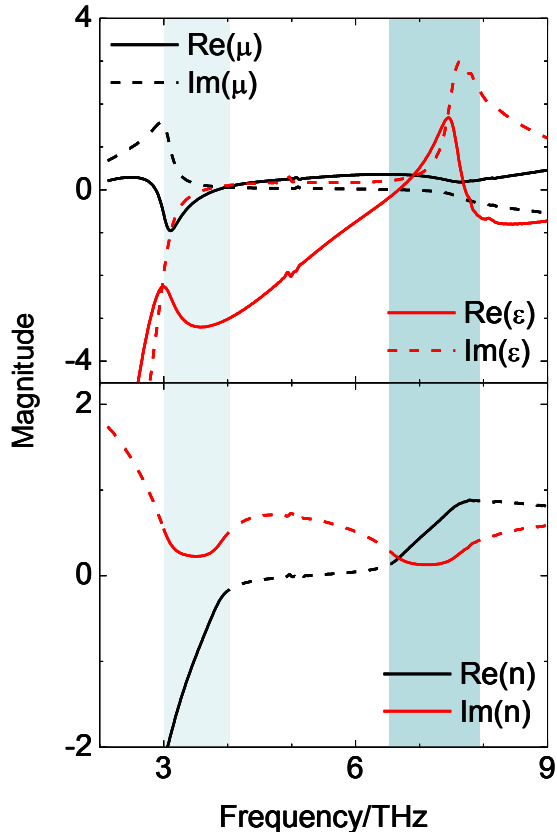

Fig. 3: Spectral dependence of materials parameters of meta-foils. Retrieval calculation of the relative complex permittivity ε, permeability μ, and refractive index n of the 1SE sample. Left-handed and right-handed pass-bands are highlighted at 3-4 THz and 6.5-8 THz, respectively.

When the incidence angle $\alpha$ is varied such that the magnetic field component normal to the induction loops changes while the electric field points along the S-strings independently of $\alpha$, the magnetically excited left-handed peak at 3.2 THz should vary as $\cos\alpha$ (Fig. 4) which is the case over the wide range from 0° to 81° implying a large latitude for choosing the incidence angle which may practically lie between ±30° or more. The relative peak width of about 25% also facilitates operation. The $\cos^2\alpha$-dependency of the electrical resonance peak is due to the relative phase shift of the electric field and the short-circuit current generated between adjacent strings. Simulated data also fit well.

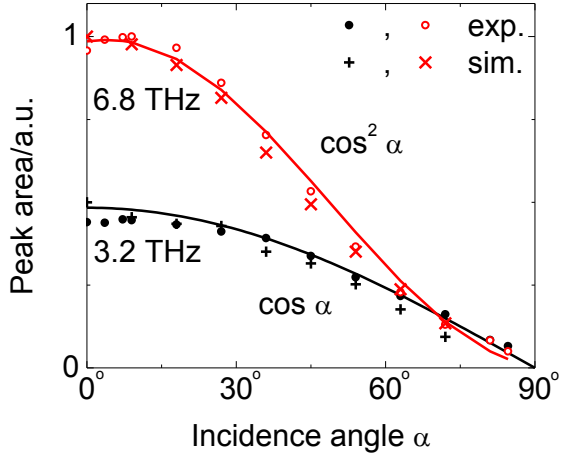

Fig. 4: Peak transmission versus incidence angle. Angular dependence of the transmission peak areas with a $\cos\alpha$-fit to the 3.2 THz left-handed magnetic resonance and a $\cos^2\alpha$–fit to the 6.8 THz right-handed electric resonance.

Meta-foils can be strongly bent without compromising their function as shown by spectra of a 2SE sample with bending radii varying from infinity to 1 cm (Fig. 5 top). The absence of any significant change of the spectra confirms the local rigidity and global flexibility of the meta-foil. Bending to a cylinder around the z-axis deforms cells from their original parallel to a wedge shape. With 1 cm bending radius, the wedge angle between two adjacent strings is 1 mrad and the change of capacitor gap ±15 nm. The relative change being only 0.003, the gap expansion in the upper half of the capacitor and the contraction in the lower cancel each other leaving no measurable effect. The slight amplitude reductions result from the decrease of the normal magnetic field component towards the edges.

From the 60%-transmission of the left-handed peak at 4 THz (Fig. 5 top) we infer that the double spacing of interconnecting lines in structure 2SE as compared to 1SE (Fig. 2) favors larger resonance peaks. Moreover, the frequency shift from 3.2 to 4 THz is explained essentially by the actual resonance loops. In the 1SE case, loops are formed by one capacitor short-circuited over the interconnecting line. In the 2SE case, half of the loops are like 1SE, but the other half consists of the canonical S-string loop that has two capacitors in series. Therefore, the frequency is 1.41 times higher, i.e., 4.5 THz. Superposition of peaks at 3.2 THz and 4.5 THz gives one peak at 3.85 THz, close to the measured 4 THz. A more accurate analytical calculation yields a frequency of $3.2\sqrt{1+\sqrt{2}/2}$ THz = 4.18 THz, in fair agreement as well.

The resonance frequency of all-metal meta-foils can be shifted by adding dielectrics which enables sensing and tuning. Peak shift and broadening caused by PMMA (Fig. 5 bottom) reflect quantitatively the relative permittivity of PMMA. The ratio of the three peaks without and with PMMA is 1.54, 1.52, and 1.5 for the 3.5, 6.3, and 7.7 THz peaks, respectively, proportional to the square root of the permittivity and, hence, to the refractive index. In fair agreement, the latter is found as 1.57 in the THz region [20]. The observed damping indicates a PMMA-induced loss. Finally, we note without showing graphs that the meta-foil was operated from room temperature to >120 °C without a detectable change of results.

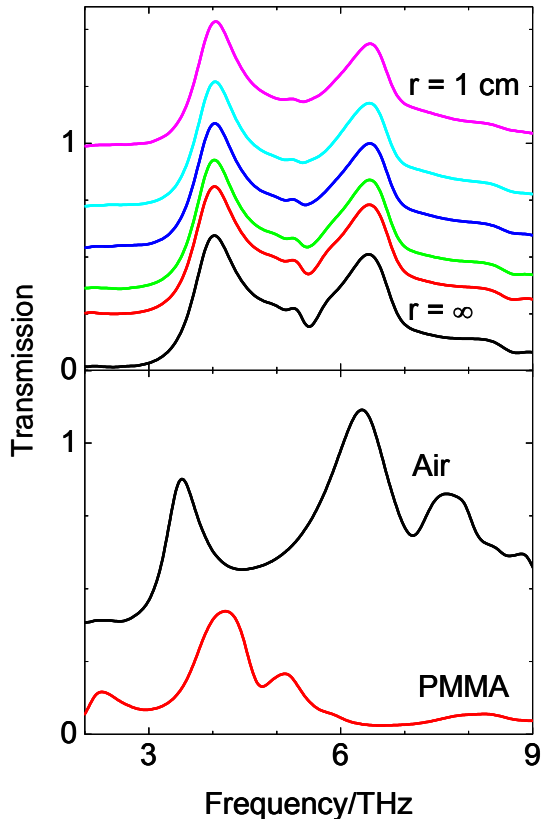

Fig. 5: Influence of mechanical deformation and dielectric environment on transmission. Spectral response under bending from flat to 1 cm radius around z axis (2SE) under normal incidence (top). Peak shift and damping upon filling with polymethylmethacrylate (PMMA) (1SP) under normal incidence (bottom).

The meta-foil is a new photonic material implementing the "solid state" of the "atoms" of the meta-material, i.e., the S-string resonant loops. Its easy handling enables a wide range of THz frequency applications. Optical elements like filters, polarizers, reflectors, and absorbers may be made from meta-foils in various geometrical shapes benefiting from the relative insensitivity of meta-foils to incidence angle, deformation, spatial or angular misalignment, and temperature. Meta-foils can be filled with dielectrics for sensing and tuning. Forthcoming work will focus on stacking to provide extended three-dimensional arrangements and enhance optical properties including resonance widths. Stacks of meta-foils may also serve as polarizing devices. Ultimately, meta-foils will allow introducing an almost arbitrary spatial distribution of refractive index by corresponding changes of geometric parameters, thus enabling index-gradient optics and radically new optical components.

**Methods**

For the present work, meta-foils have been manufactured via a 3D microfabrication process which involves three levels of photolithography and gold electroplating. Precise alignment of actual mask and already processed substrate is mandatory. The repeated gold electroplating needs a precise thickness control. The free-standing foil of metamaterial is finally released from the substrate by wet etching of a chromium sacrificial layer. Referring to Fig. 1 a), three optical masks with design patterns and alignment marks are used in our 3D multi-layer microfabrication process. Mask 1, 2, 3 are to pattern the bottom, middle, and top layers, respectively. The whole process is done on a 4 inch Si wafer as a substrate and includes a total of 24 fabrication steps.

The fabrication of the bottom layer involves 6 steps: (1) clean a Si wafer; (2) sputter deposit a Cr/Au (100 nm/50 nm) plating seed layer; (3) apply AZ resist on substrate by spin coating to a thickness of 5 μm (AZ9260, 3500 rpm/35 s) and soft bake (95°C/3 min); (4) expose to UV light under a Karl Suss Mask & Bond Aligner (MA8/BA6, dose 550 mJ/cm$^2$); (5) remove exposed AZ resist by means of the AZ 400k developer; (6) deposit gold by electroplating into the so patterned AZ resist which serves as a mould for the bottom-layer S-string parts. A pulsed current source is used for plating to achieve a high quality of the gold deposit.

The fabrication of the middle layer also involves 6 steps: (1) Hard bake the sample which includes the patterned AZ resist and gold electroplated bottom layer (100°C/30 min); (2) sputter deposit an Au (100nm) film on top of the bottom layer; (3) apply AZ resist on top of Au film by spin coating to a thickness of 5 μm (AZ9260, 3500 rpm/35 s), and soft bake (95°C/3 min); (4) precisely align sample and mask 2, and expose to UV light under the same conditions as above; (5) remove exposed AZ resist by means of the AZ 400k developer; (6) deposit gold by pulsed electroplating into the so patterned AZ resist which serves as a mould for the middle-layer S-string parts and interconnecting lines.

The fabrication of the top layer involves the same steps as for the middle layer except that mask 3 is used in step (4).

To finally achieve the free-standing and matrix-free all-metal meta-foil, the following process is carried out: (1) remove the AZ9260 resist of the top layer with AZ(R) 400T photoresist stripper; (2) remove the gold plating base of the top layer by gold etchant; (3) repeat steps (1) and (2) for middle and bottom layers; (4) remove the Cr layer of plating seed layer (Cr/Au) by Cr etchant. As this Cr layer acts as a sacrificial layer, the meta-foil is released from the Si substrate.

**Acknowledgements**

Work partly performed at SSLS under NUS Core Support C-380-003-003-001, A*STAR/MOE RP 3979908M and A*STAR 12 105 0038 grants. HSC, XXC, and BIW want to acknowledge the support from NNSFC (Nos. 60801005, 60531020, 60990320 and 60990322), the FANEDD (No. 200950), the ZJNSF (No. R1080320), the MEC (No. 200803351025), the ONR (No. N00014-06-1-0001), and the DAF (No. FA8721-05-C-0002).